\documentclass[aps,prd,nofootinbib,showpacs,superscriptaddress, preprint]{revtex4-1}
\pdfoutput=1
\usepackage[utf8]{inputenc}
\usepackage{amsmath,amssymb}
\usepackage{epsfig}
\usepackage{graphicx}
\usepackage[usenames,dvipsnames]{color}
\usepackage[colorlinks,citecolor=blue]{hyperref}
\usepackage{color}
\usepackage{subfigure}

\begin{document}
\title{Inflation and Reheating with a Fermionic Field}

\author{Abhass Kumar}
\email{abhasskumar@hri.res.in}
\affiliation{Harish-Chandra Research Institute, Chhatnag Road, Jhunsi, Allahabad 211019, India}
\affiliation{Homi Bhabha National Institute, Training School Complex, Anushakti Nagar, Mumbai 400094, India}

\begin{abstract}
Inflation has long been the accepted paradigm for understanding the early universe. Most models of inflation have a scalar field acting as the inflaton particle which decays after inflation during a process called reheating into standard model (SM) particles. In this work, we consider a fermion as the inflaton field. Noether symmetry arguments show that an exponentially expanding universe is possible with a fermion if it is coupled non-minimally to gravity. We use such a scenario in this model to study both inflation and reheating. We find relations between the various parameters involved in this model which include the non-minimal coupling strength $\xi$, the mass $m$ and the Yukawa coupling $Y$.
\end{abstract}
\maketitle

\section{Introduction}
The last few decades have been very wonderful for the physics of the early universe, specially with regards to inflation \cite{PhysRevD.23.347, 1982PhLB..108..389L}. Inflation has for long been the accepted paradigm for the early universe and a very observationally sound explanation for the homogeneity and isotropy of the universe though CMB studies \cite{Akrami:2018odb,Komatsu:2010fb}. Although, scalar inflaton (including Higgs boson as the inflaton) \cite{Bezrukov:2007ep,Bezrukov:2010jz,PhysRevLett.80.1582} has been the dominant model for a long time, people have also studied the fermionic field as the inflaton \cite{Saha:1996bx,PhysRevD.74.124030,Ribas:2007qm,PhysRevD.72.123502,Ribas:2016ulz,1742-6596-306-1-012052,ArmendarizPicon:2003qk,PhysRevD.77.123535,Alexander:2014eva}. In \cite{deSouza:2008az}, Noether symmetry was used to show that a non-minimal coupling of the fermion field with gravity is required to have an exponentially expanding phase of the universe. 

After inflation ends, the universe expands by such a huge amount that anything present before inflation gets diluted to negligible levels. The universe needs to be repopulated by the particles of the standard model. This process is called reheating \cite{1982PhLB..108..389L} and it occurs while the inflaton oscillates in a potential that can be approximated by a simple harmonic one. As the inflaton oscillates, it starts decaying into other particles which are relativistic. These particles repopulate and reheat the universe.

In this work, we use a fermionic field coupled non-minimally to gravity as the inflaton. We calculate the value of the non-minimal coupling as a function of mass and the Yukawa coupling to produce successful reheating of the universe after inflation. We also calculate the radiation energy density produced during reheating and the reheating temperature as functions of mass and the Yukawa coupling. In section 2, we write down the action of the model. Section 3 is devoted to achieving an exponentially expanding universe while reheating is studied in some detail in section 4. We follow it all up by the conclusion and discussions in section 5.

\section{The inflationary action}
The action for a fermion field coupled non-minimally to gravity is given as follows:
\begin{align}
S=\int d^4x \sqrt{-g}\left[F(\Psi)R+\frac{i}{2}(\bar\psi \tilde\gamma^\mu D_\mu\psi-(D_\mu\bar\psi)\tilde\gamma^\mu\psi)-V  \right],
\end{align}
where $F(\Psi)$ is a function of the scalar bilinear $\Psi=\bar\psi\psi$. $V$ is the fermionic potential
\begin{equation}
V=V_0+V_{int},
\end{equation}
with $V_0=m\bar\psi\psi$ and $V_{int}=Y\bar\psi H L+h.c.$. $\tilde\gamma^\mu$ are the generalized gamma matrices given by $\tilde\gamma^\mu=e^\mu_\nu\gamma^\nu$ where $e^\mu_\nu$ are the tetrad fields. $D_\mu$ are the covariant derivative operators given by:
\begin{eqnarray}
D_\mu\psi&=&\partial_\mu\psi-\Omega_\mu\psi,\\
D_\mu\bar{\psi}&=&\partial_\mu\bar{\psi}+\bar{\psi}\Omega_\mu,\\
\Omega_\mu&=&-\frac{1}{4}g_{\rho\sigma}\left[\Gamma^\rho_{\mu\delta}-e^\rho_b(\partial_\mu e^b_\delta)\right]\tilde\gamma^\delta\tilde\gamma^\sigma.
\end{eqnarray}
$\Omega_\mu$ is called the spin connection while $\Gamma^\mu_{\nu\delta}$ are the Christoffel symbols

The FRLW metric is given by $d\tau^2=dt^2-a(t)^2(dx^2+dy^2+dz^2)$ following the $(+,-,-,-)$ sign convention with $a(t)$ being the scale factor. The various generalized gamma matrices and the spin connections for this metric are:
\begin{align}
\tilde{\gamma}^0=\gamma^0,\;\;\tilde{\gamma}^i=\frac{\gamma^i}{a(t)};\;\;\;\;\Omega_0=0,\;\;\Omega_i=\frac{1}{2}\dot{a}^2
\gamma^i\gamma^0.
\end{align}

In \cite{deSouza:2008az,Grams:2014woa}, it was shown by Noether symmetry arguments that an exponentially expanding cosmological solution can be obtained if $F'\neq 0$ where the $'$ denotes a derivative wrt the bilinear $\Psi=\bar{\psi} \psi$. We take $F(\Psi)=\frac{1}{2}(M^2-\xi\Psi)$ where $M^2=\frac{1}{8\pi G}$ is the reduced Planck mass.

\section{Inflation}
During inflation, all other fields apart from the inflaton give zero contribution despite being present in the Lagrangian. Any contribution due to them gets diluted to negligible amounts and we can ignore such fields. Therefore we can ignore the $V_{int}$ term from the Lagrangian. The remaining terms give the following equations of motion for $\psi$ and $\bar{\psi}$ by minimizing the action:
\begin{eqnarray}
\dot{\bar{\psi}}+\frac{3}{2}H\bar\psi-iV_0'\bar\psi\gamma^0+6iF'\bar\psi\gamma^0(\dot H+2H^2)&=&0,\label{psibareq}\\
\dot{\psi}+\frac{3}{2}H\psi-iV_0'\gamma^0\psi-6iF'\gamma^0\psi(\dot H+2H^2)&=&0,\label{psieq}
\end{eqnarray}
where $'$ denotes a derivative with respect to the bilinear $\Psi$ and $H=\frac{\dot{a}}{a}$ is the Hubble parameter. Multiplying Eq. \ref{psibareq} by $\psi$ from the right and adding it to Eq. \ref{psieq} multiplied by $\bar\psi$ from the left gives us:
\begin{equation}
\frac{d\Psi}{dt}+3H\Psi=0.\label{PsiEq}
\end{equation}
which is the equation of motion for $\Psi$ during inflation. Substituting for $H$, we obtain a corresponding equation for $\Psi$ as a function of $a$:
\begin{eqnarray}
\frac{d\Psi}{dt}+3\frac{da}{dt}\frac{\Psi}{a}&=&0,\nonumber\\
\frac{d\Psi}{da}+3\frac{\Psi}{a}&=&0,\\
\Psi&=&\frac{\Psi_0}{a^3},\label{Psivsa}
\end{eqnarray}
where $\Psi_0$ is the initial value of $\Psi$.

The energy density of the fermion field is:
\begin{equation}
\rho_f=V-6HF^\prime\dot\Psi,
\end{equation}
which can be used in the Friedmann equation to get:
\begin{equation}
H^2=\frac{V-6HF^\prime\dot\Psi}{6F}.\label{Hub}
\end{equation}
Using Eq. \ref{PsiEq} in Eq. \ref{Hub}, we obtain:
\begin{equation}
H^2=\frac{m\Psi}{3M^2+6\xi\Psi}.
\end{equation}

An exponentially expanding universe can be obtained if $H$ is almost a constant. This can be satisfied by keeping $\xi\Psi\gg M^2$ in which case, we can neglect $M^2$ from the denominator to get 
\begin{equation}
H^2_I=\frac{m}{6\xi}.
\end{equation}
The subscript $I$ is used to denote the inflationary era. On the other hand when $\xi\Psi\ll M^2$, we recover a matter dominated universe.
Looking at Eq. \ref{Psivsa}, we can gauge an estimate for $\xi\Psi$ such that it is much larger compared to $M^2$ during inflation while becoming much smaller than $M^2$ so that the non-minimal coupling term is diluted and can be neglected to return the universe in a matter dominated era. This is when reheating kicks in. In the next section, while studying the reheating process, we will take $N=60$ as the number of $e$-foldings as the amount by which the universe expands during inflation i.e. $a$ at the end of inflation is $e^{60}$ times the $a$ at the beginning of inflation.

\section{Reheating}
Let us first define some parameters: $t_0$ is the time when inflation ends and reheating starts, $\rho_0=m\Psi_e$ is the energy density in the fermion field left at the beginning of reheating or the end of inflation, $\rho_\gamma$ is the energy density of relativistic particles produced as a result of the decay of the fermion field $\psi$.

Using $N=H_I t_0=60$ (considering the start of inflation as time $t=0$), we obtain $t_0$ as:
\begin{equation}
t_0=60\sqrt{\frac{6\xi}{m}}.
\end{equation}

The decay rate for the fermion field is given as:
\begin{equation}
\Gamma=\frac{mY^*Y}{8\pi},
\end{equation}
assuming that reheating occurs much before electroweak symmetry breaking so that all the SM particles into which $\psi$ decays are massless.

The inflaton will decay into relativistic SM particles efficiently only when the decay rate $\Gamma$ gets comparable to the Hubble rate during reheating $\Gamma\gtrsim H$. In the early stages of reheating the Hubble parameter goes as $H=\frac{2}{3t}$ as the universe is matter dominated. Decay of $\psi$ will start at the beginning of reheating i.e. at $t_0$ only if:
\begin{eqnarray}
\Gamma&\gtrsim &\frac{2}{3t_0},\nonumber\\
\Rightarrow \xi^{1/2}&=&\frac{4\pi}{45\sqrt{6 m}Y^*Y}.
\end{eqnarray}

\begin{figure}[h!]
\includegraphics[scale=1]{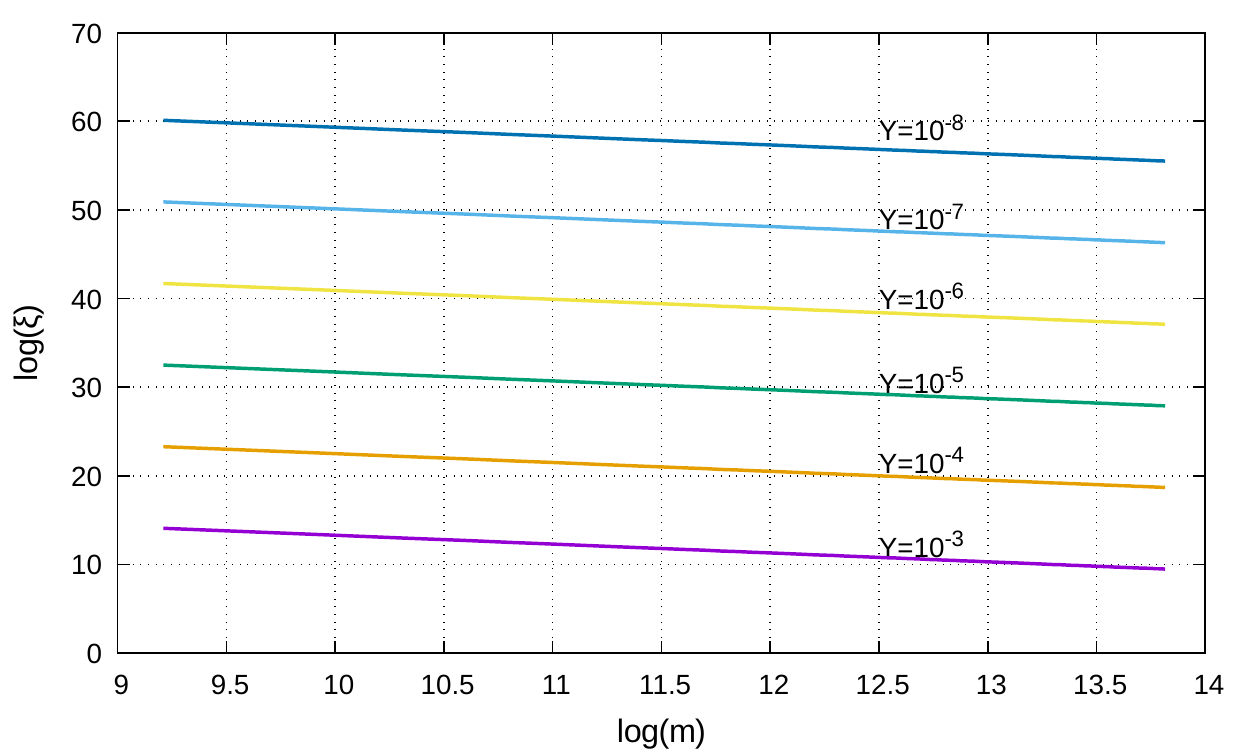}
\caption{\textit{Variation of $\xi$ as a function of the Yukawa coupling and $m$}}
\label{Fig:xivsm}
\end{figure}

The variation of $\xi$ as a function of $m$ and $Y$ is shown in Fig. \ref{Fig:xivsm} from which it is clear that $\xi$ is maximized  by minimizing both the mass $m$ and the Yukawa strength $Y$.

Since $\psi$ oscillates in a quadratic potential of $m\bar{\psi}\psi$, it evolves as a pressure-less matter fluid. The energy density $\rho_f$ stored in $\psi$ at any time decreases as $a^{-3}$. Apart from the expanding universe, $\rho_f$ also decreases exponentially due to decay of $\psi$ particles.

Once $\xi$ is known, we can take an ansatz for $\Psi_e$ such that $\xi \Psi_e \ll M^2$. This choice should also be consistent with $\xi\Psi_0\gg M^2$ during inflation when $\Psi_0\approx \Psi_e e^{60}$.

As the inflaton decays, it produces relativistic particles with energy density $\rho_\gamma$. Using the conservation of stress-energy tensor, we know that:
\begin{equation}
\frac{d\rho_{total}a^3}{dt}=-3\dot{a}a^2p_{total},\label{enerCons}
\end{equation}
where $\rho_{total}=\rho_f+\rho_\gamma$ and $p_{total}=p_\gamma=\frac{\rho_\gamma}{3}$. We also know that $\rho_f=\rho_0\left(\frac{a(t_0)}{a(t)}\right)^3\,e^{-\Gamma(t-t_0)}$. Substituting these in Eq. \ref{enerCons}, we get a differential equation for the energy density stored in radiation as follows:

\begin{eqnarray}
\frac{d(\rho_\gamma a^4)}{dt}&=&\Gamma\rho_f a^4=\Gamma\rho_0\frac{a(t_0)^3}{a(t)^3}a(t)^4e^{-\Gamma t}=\Gamma \rho_0a(t_0)^3a(t)\,e^{-\Gamma (t-t_0)}.\label{reheatEq}
\end{eqnarray}

For a matter dominated universe, $a(t)\propto t^{2/3}$ while for a radiation dominated universe, $a(t)\propto t^{1/2}$. The reheating era does start with a matter dominated universe but soon enough, the amount of radiation exceeds the amount of matter and the universe evolves accordingly. This happens during reheating itself. Therefore to solve for $\rho_\gamma$ from Eq. \ref{reheatEq}, we use $a(t)\propto t^w$. If $w$ is a constant, we have:
\begin{equation}
\rho_\gamma = \frac{\rho_0 \,e^{\Gamma t_0}\,t_0^{3w}}{\Gamma^w\; t^{4w}}\Gamma(w+1,\Gamma t_0,\Gamma t),\label{reheat}
\end{equation}
where $\Gamma(a,b,c)$ is the generalized incomplete Gamma function. As it turn out, $w$ is a constant for a large duration of the reheating phase, first as $w=\frac{2}{3}$ followed by a short transition phase after which $w=\frac{1}{2}$ for the remainder of reheating which allows us to use Eq. \ref{reheat}.

\begin{figure}[h!]
\centering
\includegraphics[scale=1]{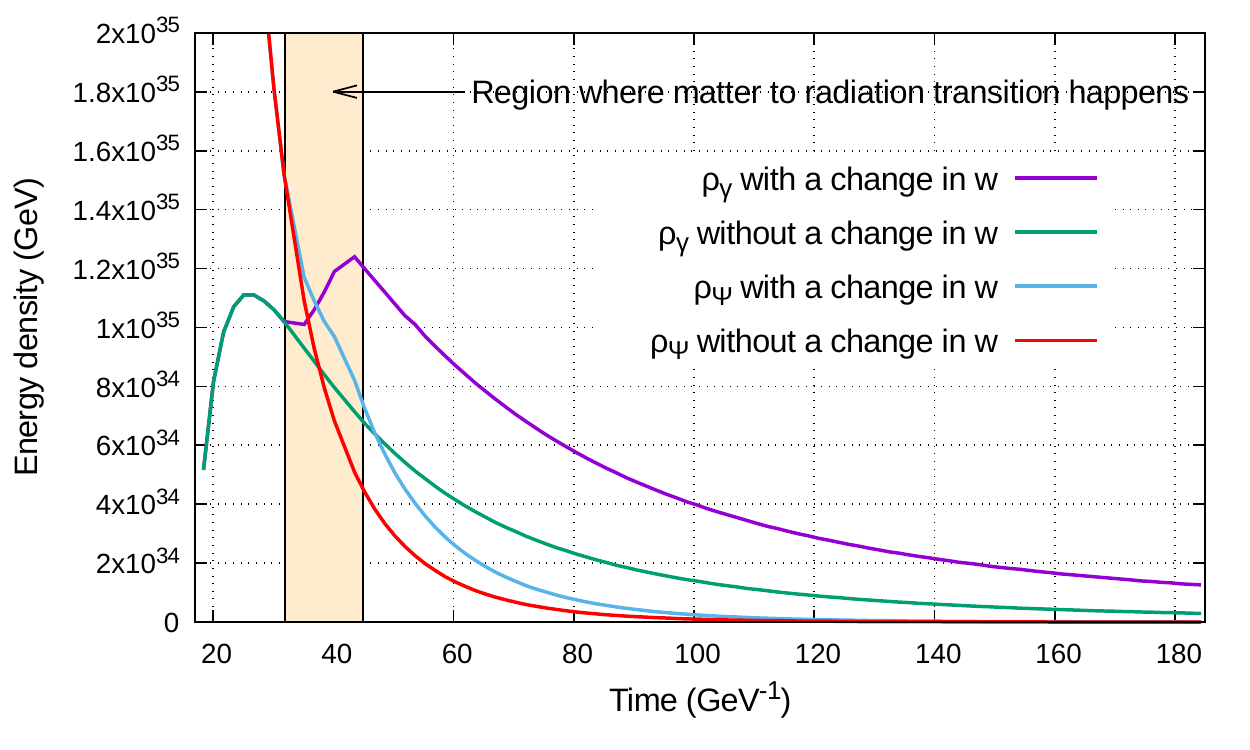}
\caption{\textit{The figure shows the energy densities stored in radiation and the inflaton as time passes during reheating. The green curve for radiation energy density and the orange curve for inflaton energy density have $w=2/3$ which is true for matter dominated universe. However, the blue curve and the purple curve have kinks which is where $w$ starts changing from $2/3$ to $1/2$ to reflect a shift from matter domination to radiation dominated era. The parameter values are: $m=10^5$ GeV, $Y^*Y=10^{-5}$, $\xi=1300$, $\Psi=10^{31}$ at the beginning of reheating}}
\label{Fig:energy}
\end{figure}

\begin{figure}[h!]
\centering
\includegraphics[scale=1]{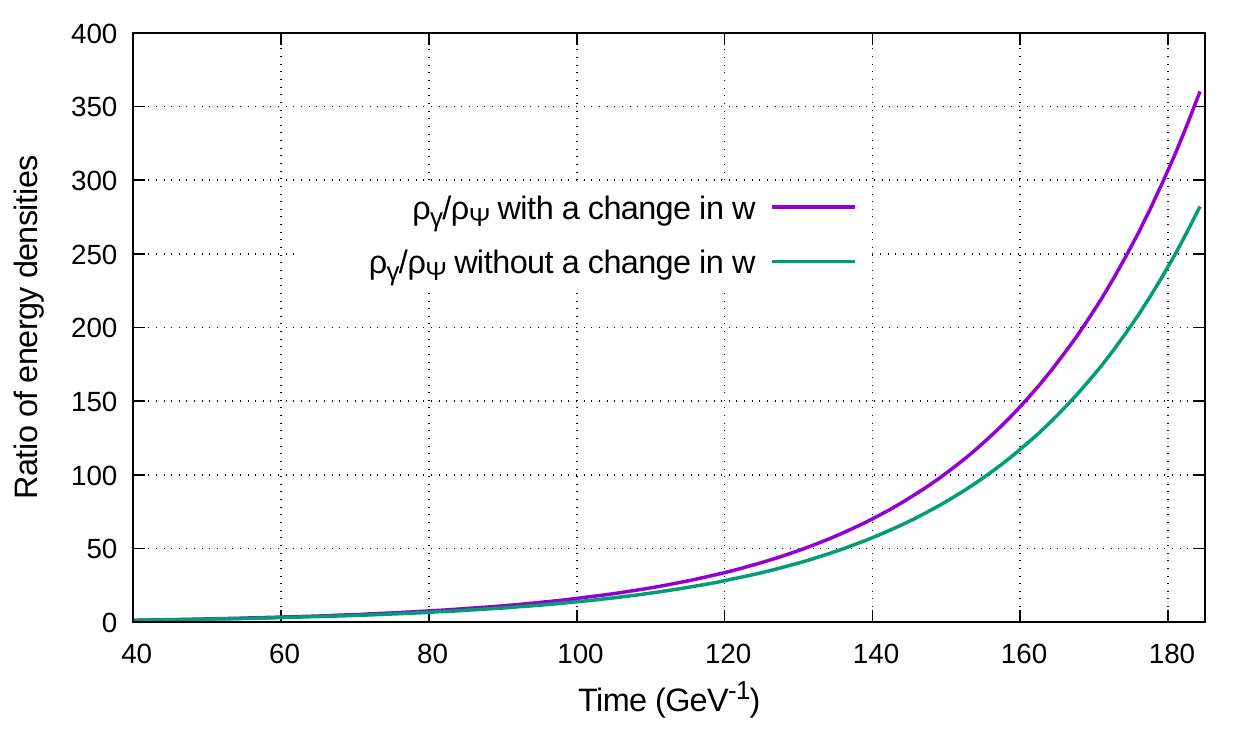}
\caption{\textit{The ratio of the energy density in radiation to the energy density in inflaton is shown here. The parameters are the same as in Fig. \ref{Fig:energy}}}
\label{Fig:ratio}
\end{figure}

\begin{figure}[h!]
\centering
\includegraphics[scale=1]{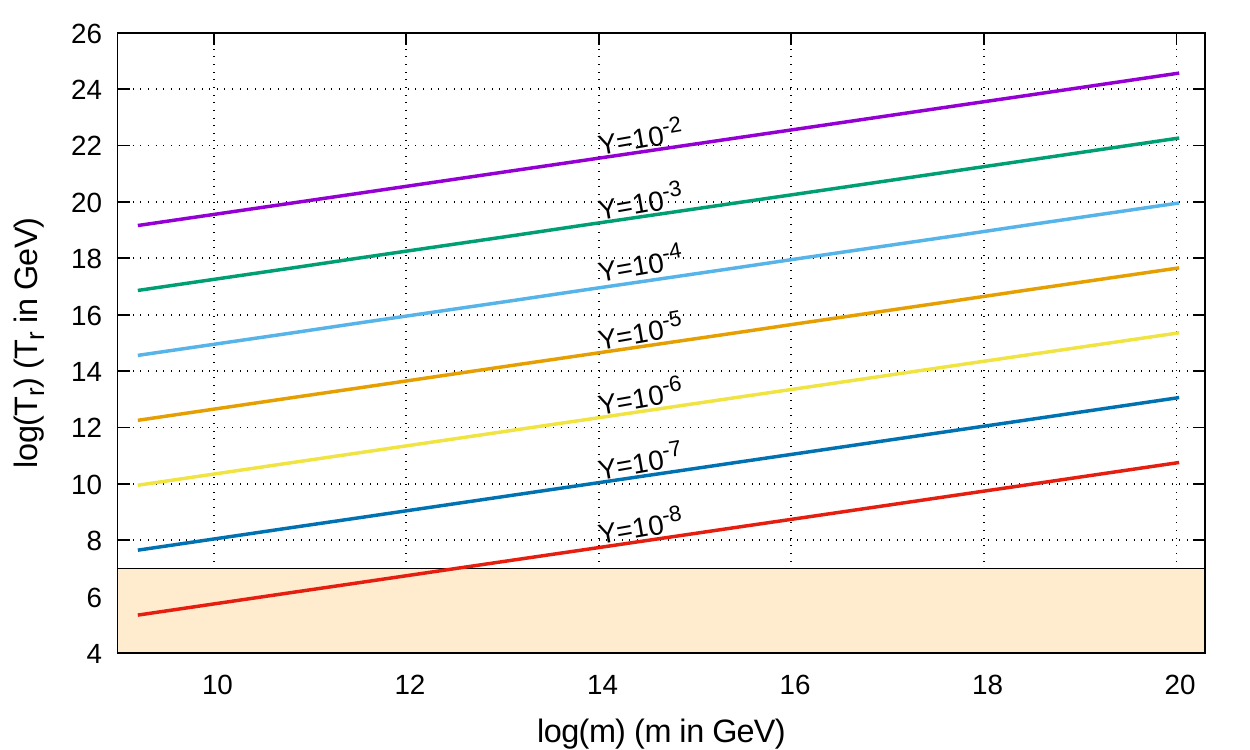}
\caption{\textit{Variation of the reheating temperature with $m$ and $Y$. Highest reheating temperatures are obtained for higher mass and higher Yukawa coupling strength. The lower shaded region is the excluded region if reheating occurs much above the electroweak scale. In this figure, $\Psi_e$ has been parametrized as $\xi\Psi_e=10^{-3}\,M^2$}}
\label{Fig:reheat}
\end{figure}

We now calculate the reheating temperature $T_r$ given by:
\begin{equation}
T_r=\left(\frac{30\rho_\gamma}{\pi^2\,g_*}\right)^{1/4},
\end{equation}
where $g_*$ is the number of degrees of freedom in the relativistic plasma. For SM, it is 109. The reheating temperature shows a wide range from hundreds of GeV to $10^{10}$ GeV depending on the mass and the Yukawa coupling choice. The variation of $T_r$ is shown in Fig. \ref{Fig:reheat}. We can see that the highest reheating temperatures of $O(10^{10})$ GeV are obtained for high mass and large Yukawa values while the lowest reheating temperatures of the order of the weak scale $\simeq 10^2$ GeV occur for low mass and low Yukawa couplings. Since we started with the assumption that reheating occurs much above electroweak breaking scale, all the parameter space with low mass and low Yukawa values is excluded.

\section{Conclusion and discussions}
In this work, we have studied inflation and reheating using a fermionic field coupled non-minimally to gravity. We calculated the value of the non-minimal coupling $\xi$ required to reheat the universe at the end of inflation. We also calculated the amount of relativistic energy density produced by decay of the inflaton and made an estimate of the effect produced by a transition from matter dominated universe in the beginning of reheating to a radiation dominated universe. At the end, we calculated the reheating temperature as a function of the mass and the Yukawa coupling.

Possibilities of $\psi$ being a freeze-in dark matter candidate can be explored in this scenario in the regions where the mass $m$ is small compared to $T_r$ such that there is enough time for other particles to produce $\psi$ particles before it becomes kinematically constrained. Such an analysis would require adding an extra symmetry to stabilize the $\psi$ particles with possible Higgs mediated interactions as a direct detection channel. This would combine fermionic inflation with dark matter phenomenology. In the case that the fermion $\psi$ is a Majorana particle, it can be a right handed neutrino, allowing us to combine neutrino masses with inflationary physics, while still being a candidate for freeze-in dark matter.

\vskip 1cm
{\bf Acknowledgments}\\
The author would like to thank the Department of Atomic Energy (DAE) Neutrino Project under the XII plan of Harish-Chandra Research Institute. This work was supported in part by the INFOSYS scholarship for senior students.

\bibliography{refsI.bib}
\end{document}